\documentclass[conference]{IEEEtran}
\IEEEoverridecommandlockouts
\usepackage{mathrsfs}
\usepackage{amsfonts}
\usepackage{ifpdf}
\usepackage{cite}
\ifCLASSINFOpdf
\usepackage[pdftex]{graphicx}
\else \fi
\usepackage{amsmath}
\usepackage{array}
\usepackage{mdwmath}
\usepackage{mdwtab}
\usepackage{eqparbox}
\usepackage[tight,footnotesize]{subfigure}
\usepackage{algorithmic}
\usepackage{algorithm}
\usepackage{amsmath}
\usepackage{epsfig}
\usepackage{ae}
\usepackage{amssymb}
\usepackage{times}
\usepackage{amsmath}   
\usepackage{booktabs}  
\usepackage{threeparttable} 
\usepackage{multirow}       
\usepackage{xcolor}

\usepackage{amsmath,amssymb,amsfonts}
\usepackage{graphicx}
\usepackage{textcomp}
\usepackage{xcolor}

\def\BibTeX{{\rm B\kern-.05em{\sc i\kern-.025em b}\kern-.08em
    T\kern-.1667em\lower.7ex\hbox{E}\kern-.125emX}}

\begin{document}

\title{AoI and Energy Consumption Oriented Dynamic Status Updating in Caching Enabled IoT Networks}


\author{\IEEEauthorblockN{Chao Xu$^{\dag,\S}$, Xijun Wang$^{*,{\diamondsuit}}$, Howard H. Yang$^{\ddag}$, Hongguang Sun$^{\dag,\S}$, and Tony Q. S. Quek$^{\ddag}$}
\IEEEauthorblockA{$^{\dag}$School of Information Engineering, Northwest A\&F University, Yangling, Shaanxi, China\\
$^{\S}$Key Laboratory for Agricultural Internet of Things, Ministry of Agriculture and Rural Affair, Yangling, China\\
$^*$School of Electronics and Communication Engineering, Sun Yat-sen University, Guangzhou, China\\
$^{\diamondsuit}$Key Laboratory of Wireless Sensor Network \& Communication,\\
Shanghai Institute of Microsystem and Information Technology, Chinese Academy of Sciences, Shanghai, China\\
$^{\ddag}$Information System Technology and Design Pillar, Singapore University of Technology and Design, Singapore}

\thanks{
This paper is supported by National Natural Science Foundation of China (61701372 and 61701372), Talents Special Foundation of Northwest A \& F University (Z111021801 and Z111021801), Key Research and Development Program of Shaanxi (2019ZDLNY07-02-01), Fundamental Research Funds for the Central Universities of China(19lgpy79), and Research Fund of the Key Laboratory of Wireless Sensor Network \& Communication (20190912).
}

}

%
%
%
%
%
%

\maketitle
\begin{abstract}
Caching has been regarded as a promising technique to alleviate energy consumption of sensors in Internet of Things (IoT) networks by responding to users' requests with the data packets stored in the edge caching node (ECN). For real-time applications in caching enabled IoT networks, it is essential to develop dynamic status update strategies to strike a balance between the information freshness experienced by users and energy consumed by the sensor, which, however, is not well addressed. In this paper, we first depict the evolution of information freshness, in terms of age of information (AoI), at each user. Then, we formulate a dynamic status update optimization problem to minimize the expectation of a long term accumulative cost, which jointly considers the users' AoI and sensor's energy consumption. To solve this problem, a Markov Decision Process (MDP) is formulated to cast the status updating procedure, and a model-free reinforcement learning algorithm is proposed, with which the challenge brought by the unknown of the formulated MDP's dynamics can be addressed. Finally, simulations are conducted to validate the convergence of our proposed algorithm and its effectiveness compared with the zero-wait baseline policy.
\end{abstract}

\section{Introduction}

Acting as a critical and integrated infrastructure, Internet of Things (IoT) enables ubiquitous connections for billions of things in our physical world, ranging from tiny, resource-constrained sensors to more powerful smart phones and networked vehicles\cite{Survey_IoT_Applications_2015}. In general, the sensors are powered by batteries with limited capacities rather than a fixed power supply. Thus, to exploit the benefits promised by IoT networks, it is essential to well address the energy consumption issue faced by sensors. Recently, caching bas been proposed as a promising solution to lower the energy consumption of sensors by reducing the frequency of environmental sensing and data transmissions \cite{IoT_Caching_2016_Network,Caching_IoT_EH_ICC_2016,IoT_Caching_2017_RL}. Particularly, by caching the data packets generated by sensors into the edge caching node (ECN), e.g., access point (AP) or mobile edge server, users can retrieve the cached data directly from the ECN instead of frequently activating sensors for status sensing and data trasnmission, thereby significantly lowering their energy consumption.

Caching multimedia contents at the edge of wireless networks has been regarded as a promising technology for the 5th Generation (5G) wireless networks and hence, has been well studied in existing work \cite{Femtocell_Caching_2013,Content_Caching_2014,Our_Caching_Comm_Mag,Tang_2019}. However, compared with multimedia contents (e.g., music, video, etc.) in traditional wireless networks, the data packets in IoT networks have two distinct features: 1) The sizes of data packets generated by IoT applications are generally much smaller than those of multimedia contents. Therefore, for IoT networks, the storage capacity of each ECN is large enough to store the latest status updates generated by all the sensors. 2) For many real-time IoT applications, the staleness of information obtained by users may significantly deteriorate the accuracy and reliability of their subsequent decisions. Therefore, the main concern for edge caching enabled IoT networks is how to properly update the cached data to simultaneously \textit{lower the energy consumption} of sensors and \textit{improve the information freshness} at users.

In order to quantify the information freshness, the age of information (AoI) has been proposed, which measures the time elapsed since the latest received packet was generated from the sensor \cite{AoI_Org_2012,AoI_Survey_2017,Our_IF_2019}. Based on this metric, some recent researches \cite{Age_Updating_2017,AoI_Cache_Updating_2019} begin to design status update strategies that optimize the AoI in caching enabled IoT networks. Specifically, authors in \cite{Age_Updating_2017} proposed a status update algorithm to minimize the popularity-weighted average of AoI values, each of which is associated to one sensor, at the cache, where the users' popularity for each sensor is assumed to be known. Study \cite{AoI_Cache_Updating_2019} further extended \cite{Age_Updating_2017} by considering the relation between the time consumption for one update and the corresponding update interval. However, in these studies, the AoI is evaluated from the ECN's (e.g., AP's) perspective instead of from the perspective of individual users. In fact, users are the real data consumers and final decision makers and hence, it is more reasonable to evaluate and optimize the AoI at users in caching enabled IoT networks. Besides, by focusing on the AoI at users, it is possible to decrease the energy consumption of each sensor by avoiding some useless status updates and meanwhile, reduce the user experienced AoI. This is due to the fact that each user could observe the information freshness of a cached data packet, only when she has asked and received it from the ECN.


In this paper, we consider a time-slotted IoT network consisting of an ECN, a sensor and multiple users, where the ECN would store the latest data packet generated by the sensor. Here, each time slot is divided into two phases, which are termed as the data delivery phase (DDP) and status update phase (SUP), respectively. At the beginning of each time slot, the ECN obtains the data query profile of users, and responds to the requests with its local cached data during the DDP. Then, the ECN will decide whether to ask the sensor to update its status and, if any, the data update will be conducted during SUP. To strike a balance between the average AoI experienced by users and energy consumed by the sensor, we first depict the evolution of the AoI at both the ENC and each user and then, formulate a dynamic status update optimization problem to minimize the expectation of a long term accumulative cost, by jointly considering the effects from the AoI and energy consumption. To solve this problem, we cast the corresponding status updating procedure into a Markov Decision Process (MDP), and design an Energy consumption and AoI oriented dynamic data Update (EAU) algorithm to solve the originally formulated problem by resorting to reinforcement learning (RL). The proposed algorithm does not require any priori knowledge of the dynamics of the MDP, and its effectiveness is further verified via simulation results.

The organization of this paper is as follows. In Section II, the description of the system model and the formulation of the dynamic status update optimization problem are presented. In Section III, we cast the dynamic status updating procedure as an MDP and then, develop a model-free RL algorithm to solve it. Simulation results are presented to show the effectiveness of our proposed scheme in Section IV and finally, conclusions are drawn in Section V.

\section{System model and problem formulation} \label{Sec:Section 2}

\subsection{Network model}

\begin{figure} [!t]
\centering
\leavevmode \epsfxsize=3.0in  \epsfbox{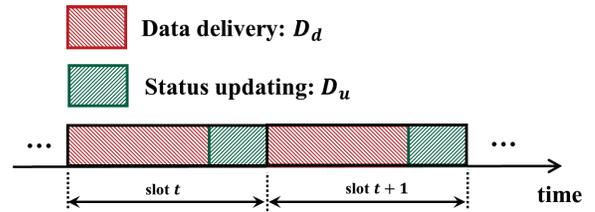}
\centering \caption{Illustration of the time slot structure.} \label{Fig:Fig1}
\end{figure}

We consider an IoT network consisting of $N$ users, one ECN, and one sensor, which serves users with time sensitive information. The user set is denoted by $\mathcal N = \{1, 2, \cdots, N\}$. We assume a time-slotted system as illustrated in Fig. \ref{Fig:Fig1}, where each slot is divided into two phases, i.e., the DDP and SUP, whose time durations are fixed and denoted by $D_d$ and $D_u$, respectively. In each slot $t$, users may request the ECN for the cached data packet, and the corresponding query profile can be expressed as $\mathbf r(t) = \left( r_1(t), r_2(t), \cdots, r_N(t)\right)$, where $r_n(t) \in \{0,1\}, \forall n \in \mathcal N$, with $r_n(t)=1$ if user $n$ requests the data, and $r_n(t)=0$ otherwise. In this paper, we consider the ECN obtains the data query profile $\mathbf r(t)$ at the beginning of each time slot $t$, and it has no prior knowledge of users' request patterns which is random.

As shown in Fig. \ref{Fig:Fig1}, the ECN would first transmit the required data packet to the corresponding users during DDP. Then, the ECN has to decide whether or not ask the sensor update its status, and, if any, receives the status update during the SUP. Let $A(t) \in \{0, 1\}$ denote the ECN's update decision made in each time slot $t$, i.e., $A(t) = 1$ if the sensor is asked to sense the underlying environment and update the status, and ${A}(t) = 0$ otherwise. We consider all data packets requested by users can be successfully transmitted from the ECN. However, when each update is transmitted from the sensor to ECN, there is a potential update failure due to the limited transmission power of the sensor, which will be presented in detail in the following subsection.

\subsection{Status updating model}
During SUP in time slot $t$, if the sensor is asked to update its status, then it has to sense the underlying environment and transmit the information packet to the ECN. The energy consumed by the sensor for each environmental sensing is assumed to be constant and denoted by $E_s$. Besides, we assume the time spent on the environment sensing is negligible and hence, the transmission time of each update packet is $D_u$. The spectral bandwidth available for the data update is $B$ and the transmit power of the sensor is fixed as $p$. Accordingly, the overall energy consumption of one status update is $E = E_s + p D_u$. Let $g(t)$ denote the channel power gain from the sensor to ECN in time slot $t$, and $\delta ^2$ the AWGN power density at the ECN. It is assumed that the channel between the sensor and ECN follows the quasi-static Rayleigh fading with mean $\bar g$, i.e., its power gain remains the same over one SUP and changes independently across time slots.

The size of each update packet is considered to be constant and denoted by $F$. To complete the data transmission in one SUP, the required transmission rate is set to $R = \frac{F}{D_u}$. In this light, if the sensor is asked to transmit its update packet, a transmission failure can occur when the required SNR threshold cannot be reached. Particularly, according to Shannon's formula, to meet the transmission rate requirement $R$, the SNR threshold should be set as
\begin{align}  \label{Eq:SINR_Threshold}
{\gamma^T} =  {{2^{\frac{{{R}}}{{{B}}}}} - 1}.
\end{align}
Furthermore, the update failure probability $P_f$ for one update transmission, if there is any, can be expressed as \cite{Wireless_Comm,IoT_Updating_2019_Reyleigh}
\begin{align}  \label{Eq:Outage_Pro}
{P_f} = \mathsf P(\gamma(t) < \gamma^T) = 1- exp(-\frac{\gamma^T}{\bar \gamma})
\end{align}
where $\gamma(t)$ denotes the SNR with the corresponding transmission, and $\bar \gamma=\frac{p \bar g}{B \delta ^2}$ the average SNR.

For ease of expression, we use $Z(t) \in \{0, 1\}$ to denote whether the asked update is successfully transmitted to the ECN, i.e., $Z(t) = 1$ if the update is successfully delivered and $Z(t) = 0$ otherwise. Since the sensor will only generate and transmit the update packet when it is asked by the ECN, we further have $Z(t) \leq A(t)$ with each time slot $t$.

\subsection{Performance metric and problem formulation}

In this paper, we consider both the AoI at the users and the ECN just before the decision moment in each time slot, i.e., the beginning of the status updating phase. Before formally depicting the evolution of AoI at each user, we first give the dynamics of AoI for the stored data packet from the ECN's perspective.
Particularly, we denote the AoI at the ECN in slot $t$ by the $\Delta _0(t)$, whose evolution can be expressed as
\begin{align} \label{Eq:AoI_Cache_Evolution}
\Delta_0(t)  =
\begin{cases}
D_u + D_d &\textrm{if} \ Z_u(t-1)=1\\
\Delta_0(t-1)+ D_u + D_d, & \textrm{otherwise}
\end{cases}
\end{align}
where $Z(t-1)=1$ means in the previous slot $t-1$ an update packet was successfully delivered to the ECN.

Similarly, let $\Delta_n(t)$ denote the AoI at each user $n$ with slot $t$. Then, its dynamics can be expressed as
\begin{align} \label{Eq:AoI_Evolution_Per_User_Per_Data}
& \nonumber \Delta_n(t) \\ & \nonumber =
\begin{cases}
D_u+D_d, &\textrm{if} \ r_n(t) Z(t-1)=1\\
\Delta_0(t-1) +D_u+ D_d \! & \textrm{if} \ r_n(t)=1 \!  \& \! Z(t-1)=0\\
\Delta_n(t-1)+D_u+D_d, & \textrm{otherwise}
\end{cases} \\
& \mathop {\rm{ = }}\limits^{\left( a \right)}
\begin{cases}
\Delta_0(t),  & \textrm{if} \ r_n(t)=1\\
\Delta_n(t-1)+D_u+D_d, & \textrm{otherwise}
\end{cases}
\end{align}
where the product $r_n(t) Z(t-1)=1$ means the packet required by user $n$ was just successfully updated to the ECN in the previous slot $t-1$, and (a) holds because of Eq. (\ref{Eq:AoI_Cache_Evolution}). Without loss of generality, we initialize $\Delta_0(1)=\Delta_n(1)= D_u+ D_d, \forall n \in \mathcal N$. Based on the above analyses, we note that, even obtaining data from the same ECN, \textit{different AoIs }may be experienced by different users, which would be ignored if we just consider the dynamics of the AoI at the ECN.

To strike a balance between the average AoI experienced by users and energy consumed by sensors, we define the overall cost associated with the update decision made by the ECM in each slot $t$ as
\begin{align} \label{Eq:Totoal_Cost}
L(t) = \beta_1 \sum\nolimits_{n = 1}^N {\omega_n \Delta_n(t)} + \beta_2 A(t)E
\end{align}
where $\omega _n$ denotes the weight allocated to user $n$, i.e., $\omega _n \in (0, 1), \forall n \in \mathcal N$ and $\sum\nolimits_{n = 1}^N \omega_n = 1$, $E$ represents the energy consumed to complete one status update, and $\beta_1$ as well as $ \beta_2$ are parameters used to nondimensionalize the equation and meanwhile, to make a trade-off between the experienced AoI and consumed energy in the network.

In this work, we aim at designing a dynamic status updating strategy to minimize the expectation of a long term accumulative cost, i.e.,
\begin{eqnarray}   \label{Eq:Minimize_Cost}
\textbf{P}: &\mathop { \min}\limits_{ \mathbf {A}}& \mathop {\lim }\limits_{T \to \infty }  \mathbb E \big[ \sum\nolimits_{t = 1}^T \lambda^{t-1} L(t) \big] \\
&\textrm{s.t.} & \mathbf {A} = \left( {A}(1), {A}(2), \cdots, {A}(T)\right)
\end{eqnarray}
where $\mathbf {A}$ denotes a sequence of update decisions made by the ECN from slot $1$ to $T$, and $\lambda$ the discount rate introduced to determine the importance of the present cost and meanwhile to guarantee the long term accumulative cost finite. The formulated dynamic optimization problem will be solved in the following section.

\section{Reinforcement learning algorithm design}
In this section, we first formulate the dynamic status updating procedure as an MDP and then, develop a model-free RL algorithm, with which the challenge brought by the unknown of transition probability in the formulated MDP could be addressed.

\subsection{MDP formulation}
The concerned dynamic status updating can be formulated into an MDP consisting of a tuple $\mathcal M = \Gamma\left(\mathbb S, \mathbb A, U\left({ \cdot , \cdot }\right)\right)$, which is depicted as follows:

\begin{itemize} \setcounter{enumi}{0}
\item \textbf{State space $\mathbb S$}: At each time slot $t$, the state $\mathcal S(t)$ is defined to be the combination of the AoIs at the ECN and users just before the data updating phase, i.e., $\mathcal S(t) = \left(S_0(t), S _1(t), S _2(t), \cdots, S _N(t)\right)$, where $S _m(t) = \Delta _m(t), \forall m \in \{0\} \cup\mathcal N$. We set the maximum AoI at the ECN and users to be $\Delta_{\max} = M_{\max}D$ where $D= D_u+D_d$ and $M_{\max}$ is an integer. In this light, the state space $\mathcal S$ is finite and can be expressed as
    \begin{align} \label{Eq:State_space}
    \mathbb{S} = \{{{\mathbb S_0}} \times {{\mathbb S_1}} \times {{\mathbb S_2}} \times \cdots \times {{\mathbb S_{N}}}\}
    \end{align}
    with $\mathbb S_m = \{D, 2D, \cdots, M_{\max}D\}, {\forall m \in  \{0\} \cup\mathcal N}$.
\item \textbf{Action space $\mathbb A$}: At each time slot $t$, the ECN could ask the sensor to sense the underlying environment and update its generated data packet. Therefore, the action space can be expressed as $\mathbb A = \{0, 1\}$, where $A = 0$ ($A \in \mathbb A$) means the ECN would not ask the sensor to conduct its status update.
\item \textbf{Reward function $U\left({ \cdot , \cdot }\right)$}: The reward function is a mapping from a state-action pair $\mathcal S \times A$  to a real number, which is used to quantify the obtained reward by choosing a action $A \in \mathbb A$ at a state $\mathcal S \in \mathbb S$. In this work, at each time slot $t$, when given a state $\mathcal S(t)$ and choosing action $\mathcal A(t)$, we define the reward function as
    \begin{align} \label{Eq:Utility_function}
    \nonumber &U\left({ \mathcal S(t), A(t)}\right) \\ \nonumber
    &= C -  \left(\beta_1 \sum\nolimits_{n = 1}^N {\omega_n \left(\Delta_n(t)+ D_u\right)} + \beta_2 A(t)E\right) \\
    & \mathop  = \limits^{(a)}  \mathop  C - \beta_1 D_u - L(t) = C_1 - L(t)
    \end{align}
    where $C$ is a constant used to ensure the reward positive, and (a) follows Eq. (\ref{Eq:Totoal_Cost}). Here, $D_u$ is introduced to evaluate the effect of the chosen action after it is conducted, i.e., after the SUP is over.
\end{itemize}

For an MDP, the goal is generally to find a deterministic stationary policy $\pi: \mathbb S \to \mathbb A$ to maximize the long term expected discounted return. Particularly, a policy is said to be deterministic and stationary if: 1. Given the state, only one certain action is chosen; 2. The policy is irrelevant to time. In this work, we aim at deriving a deterministic stationary policy $\pi ^*$ that maximizes the long term expected discounted reward with the initial state $\mathcal S(0)$, i.e.,
\begin{align} \label{Eq:Optimal_Policy}
\nonumber \pi^* &= \mathop {\arg }\limits_ \pi \max \mathop {\lim }\limits_{T \to \infty }  \mathbb E \big[ \sum\nolimits_{t = 0}^T \lambda^{t} U\left({ \mathcal S(t), A(t)}\right) \left| \mathcal S(0) \right. \big] \\
& \mathop  = \limits^{(a)}  \mathop {\arg }\limits_ \pi \min \mathop {\lim }\limits_{T \to \infty }  \mathbb E \big[ \sum\nolimits_{t = 1}^T \lambda^{t-1} L(t) \big].
\end{align}
where (a) holds when all the elements in $\mathcal S(0)$ are set to $D$. Comparing Eq. (\ref{Eq:Minimize_Cost}) with Eq. (\ref{Eq:Optimal_Policy}), we note that $\pi ^*$ can also be used to derive a solution of the original Problem \textbf{P}.

As shown in Eq. (\ref{Eq:Optimal_Policy}), in each time slot $t$, the immediately obtained reward $U\left({ \mathcal S(t),  A(t)}\right)$ does affect the cumulative reward in the long run. Therefore, to find the optimal strategy $\pi ^*$, it is essential to accurately estimate the long-term effect of each decision, which is nontrivial because of the causality. In this work, we resort to RL, and design a model-free algorithm, named as EAU, to solve it, which will be presented in detail in the following subsection.

\subsection{EAU algorithm}

\begin{figure*} [!t]
\begin{align} \label{Eq:Act_Value_Fun}
Q_{\pi}(\mathcal S,  A) = \mathbb E_{\pi} \big[ \sum\nolimits_{l = 0}^\infty  {{\lambda ^{t - 1}}} U\left( {{\cal S}(t + l),{ A}(t + l)} \right)\left| {{\cal S}(t) = \mathcal S, A(t) =  A} \right. \big]
\end{align}
\hrulefill
\end{figure*}

\begin{figure*} [!t]
\begin{align} \label{Eq:Act_Value_Bellman_Opt_Fun}
Q_{\pi^*}(\mathcal S, A) &= \mathbb E_{\pi *} \big[ U\left( {\mathcal S(t),\mathcal A(t)} \right) + \gamma \mathop {\max }\limits_{ A'  \in \mathbb A} {Q_{{\pi ^*}}}({\mathcal S}(t + 1), A') \left| {{\mathcal S}(t) = \mathcal S, A(t) = A} \right. \big] \\ \nonumber
& =  U\left( {\mathcal S(t),A(t)} \right) + \gamma \sum\limits_{\mathcal S \in \mathbb S'} {\mathsf P\left( {{\mathcal S}'\left| {{\mathcal S},{ A}} \right.} \right)} \mathop {\max }\limits_{A' \in \mathbb A} {Q_{{\pi ^*}}}(\mathcal S', A')
\end{align}
\hrulefill
\end{figure*}

For each deterministic stationary policy $\pi$, we define the action-value function as shown in Eq. (\ref{Eq:Act_Value_Fun}), with $( \mathcal S, A)$ denoting the initial action-state pair, and the corresponding Bellman optimality equation can be expressed in Eq. (\ref{Eq:Act_Value_Bellman_Opt_Fun}) \cite{RL_Introduction}, where ${\mathsf P\left( {{\mathcal S}'\left| {{\mathcal S},{A}} \right.} \right)}$ denotes the transition probability from one state $\mathcal S$ to $\mathcal S'$ by conducting an action $A$. Essentially, if that probability is known, the Bellman optimality equations are actually a system of equations with $\left| {\mathbb S \times \mathbb A} \right|$ unknowns, where $\left| {\cdot} \right|$ represents the cardinality of a set. In that case, for a finite MDP, we could derive the unique solution by utilizing model based RL algorithms, e.g., dynamic programming\cite{RL_Introduction}.

However, in this work, the transition probability cannot be regarded as a prior knowledge due to the unknown of the users' request patterns and mean channel gain from the sensor to ECN. To deal with the corresponding challenge, we develop a model-free algorithm by resorting to expected Sarsa, one kind of temporal-difference (TD) RL algorithms. We note that, compared with traditional TD algorithms, e.g., classic Sarsa and Q-learning, expected Sarsa may achieve better performance with small additional computational cost for many problems \cite{RL_Introduction}. Our developed EAU algorithm is shown in Algorithm 1, where $\hat{\mathbf Q}(t)$ is a row vector with $\left| {\mathbb S \times \mathbb A} \right|$  elements, denoting the estimate of the action-value function obtained after the $t$-th iteration. In this algorithm, the loop is repeated until the number of iterations reaches the maximum value $T_{\max}$.

\begin{algorithm}
\caption{ Energy consumption and AoI dependent dynamic data Update (EAU) algorithm.}
\begin{algorithmic}[1] \label{Alg:EAU_algorithm}
\STATE \textbf{Initialization:}
\STATE Set $t=0$, $\hat{\mathbf Q}(t)= \mathbf 0$, $C_1$ with Eq. (\ref{Eq:Constant_C1}), and $\mathcal S(t) = \{D, D, \cdots, D\}$ as the initial state. Randomly select an action $A(t)$.
\STATE \textbf{Go into a loop:}
\FOR{$t < T_{\max} $}
\STATE \textbf{Action Selection:} Choose a action $A(t)$ according to the probability distribution shown in Eq. (\ref{Eq:Prob_Dis_E_Greddy}), i.e., $\epsilon$-greedy.
\STATE \textbf{Acting and observing:} Take action $A(t)$, obtain a reward $U\left({ \mathcal S(t), A(t)}\right)$, and observe a new state $ \mathcal S(t+1)$.
\STATE \textbf{$\hat{\mathbf Q}(t)$ Updating:} Update $\hat{\mathbf Q}(t+1)$ with Eq.(\ref{Eq:AV_Update})-(\ref{Eq:Estimate_Expectation}) and set $t = t+1$.
\ENDFOR
\STATE \textbf{Output:} $\hat{\mathbf Q}(t)$.
\label{Alg1:Iteration_End}
\end{algorithmic}
\end{algorithm}

At the beginning of EAU, the estimate of the action-value vector $\hat{\mathbf Q}(0)$ is initialized as a vector with all elements being 0, and all elements of the initial state $\mathcal S(0)$ are set to $D$. We note the constant $C_1$ in Eq. (\ref{Eq:Utility_function}) could be chosen arbitrarily, and, as an instance, we set it as
\begin{align} \label{Eq:Constant_C1}
C_1 =  \beta_1 \sum\nolimits_{n = 1}^N {\omega_n (M_{\max}D)} + \beta_2 E
\end{align}
with which the immediately obtained reward in each iteration $t$ is guaranteed to be positive.

When the initialization is completed, the algorithm goes into a loop. At each iteration $t$, we will first choose an action $ A(t)$ from the action space $\mathbb A$ based on the current state $\mathcal S(t)$ by looking up the action-value vector $\hat{\mathbf Q}(t)$. To balance the exploration and exploitation, we adopt the $\varepsilon$-greedy policy here, i.e., choosing an action from the space $\mathbb A$ with the following probability distribution
\begin{align}   \label{Eq:Prob_Dis_E_Greddy}
\mathsf P_{\hat{\mathbf Q}(t)} &\left( {  A(t) \left| { \mathcal S(t)} \right.} \right)\\ \nonumber
&= \begin{cases}
\varepsilon, &\text{$ A (t) = \mathop {\arg }\limits_{A \in \mathbb A} \max{ \{\hat{Q}\left({\mathcal S(t), A}\right) \in \hat{\mathbf Q}(t)\}}$} \\
1-{{\varepsilon }}, &\text{otherwise}
\end{cases},
\end{align}
where $\hat{Q}\left({\mathcal S(t), A}\right)$ denotes the estimated state-action value for the pair $\left({\mathcal S(t),  A}\right)$ with the vector $\hat{\mathbf Q}(t)$, and $\varepsilon$ belongs to $\left(0, 1\right)$.
After that, we would conduct the action $A(t)$, obtain a reward $U\left({ \mathcal S(t), A(t)}\right)$, as shown in Eq. (\ref{Eq:Utility_function}), and then observe a new state $\mathcal S(t+1)$.

Learning from the new experiences (i.e., $\mathcal S(t)$, $A(t)$ and $\mathcal S(t+1)$), we would update the estimate of the action-value vector from $\hat{\mathbf Q}(t)$ to $\hat{\mathbf Q}(t+1)$ by just updating the element associated with the pair ($\mathcal S(t)$, $ A(t)$). Particularly, if we denote the element with ($\mathcal S(t)$, $\mathcal A(t)$) in $\hat{\mathbf Q}(t)$ and that in $\hat{\mathbf Q}(t+1)$ by $\hat{Q}\left({\mathcal S(t), \mathcal A(t)}\right)$ and $\hat{Q'}\left({\mathcal S(t), \mathcal A(t)}\right)$, respectively, we have
\begin{align}   \label{Eq:AV_Update}
\hat{Q'}\left({\mathcal S(t), A(t)}\right) = \hat{Q}\left({\mathcal S(t), A(t)}\right) + \alpha \Xi(t)
\end{align}
where $\alpha \in (0, 1]$ is the learning step-size and $\Xi(t)$ denotes the obtained TD error. Here, $\Xi(t)$ can be as
\begin{align}   \label{Eq:Estimate_Error}
\Xi(t) =  &U\left({ \mathcal S(t), A(t)}\right) + \lambda \mathbb E_{\varepsilon}\big[\hat{\mathbf Q}(t) \left| \mathcal S(t+1) \right. \big] -\\ \nonumber
&\hat{Q}\left({\mathcal S(t), A(t)}\right)
\end{align}
where $\lambda$ denotes the discount rate and the expectation
\begin{align}   \label{Eq:Estimate_Expectation}
&\mathbb E_{\varepsilon}\big[\hat{\mathbf Q}(t) \left| \mathcal S(t+1) \right. \big] \\ \nonumber
&= \sum\nolimits_{A \in \mathbb A} {\mathsf P_{\hat{\mathbf Q}(t)}\left( {A \left| { \mathcal S(t)} \right.} \right) \hat{Q}\left({\mathcal S(t+1), A}\right)}
\end{align}
represents the expected action-value we will have, if we start from the state $\mathcal S(t+1)$, still evaluate the effect of actions according to the action-value vector $\hat{\mathbf Q}(t)$, and choose the action with the $\varepsilon$-greedy policy as shown in Eq. (\ref{Eq:Prob_Dis_E_Greddy}).

When the algorithm is terminated, we will obtain an estimated action-value function mapping from each state-action pair to a positive number, i.e., $\hat{\mathbf Q}(T_{\max})$. With $\hat{\mathbf Q}(T_{\max})$ and the initial state $\mathcal S = \{D, D, \cdots, D\}$, we can obtain an approximate solution of Problem \textbf{P} by keeping looking up $\hat{\mathbf Q}(T_{\max})$ and choosing the action bringing the maximum action-value in each decision time.

\section{Simulation results}

In this section, we conduct simulations to evaluate the performance of our proposed strategy. We consider an IoT network consisting of 1 sensor, 1 ECN, and 3 users, where $D_u$ and $D_d$ are both set to 1 s. To simulate the data requests of users, we assume the requests of each user arrive at the ECN according to an independent Bernoulli distribution with parameter $P_n$, i.e., in each time slot $t$, we have $\mathsf P(r_n(t)=1) = P_n, \forall n \in \mathcal N$, and $\mathsf P\left( r_n(t) = 0\right) = 1-P_n$. Here, we set $P_n = 0.6$, $\forall n \in \{1, 2, 3\}$. For the sensor, the generated data packet is 200 Kbits, the adopted transmit power $p$ is 10 mW, and  the energy consumption for sensing $E_s$ is half of that for data transmissions. The channel bandwidth is 100 KHz, the mean of channel power gain over the sensor and ECN $\bar g$ is $-120$ dB, and the AWGN power density $N_0$ is set to $-174$ dBm/Hz. Meanwhile, $M_{\max}$, the parameter related to the maximum AoI, is set to 20, and the user weight factor $\omega_n =1/3$, $\forall n \in \{1, 2, 3\}$. For our proposed algorithm EAU, the learning time $T_{\max}$ is set to $10^8$, and the learning step-size $\alpha$ is set to 0.1. Meanwhile, as in \cite{Caching_DRL_2019}, we set the discount rate $\lambda$ to $0.99$ to strength the effects of rewards obtained in the future. Besides, the number of immediate rewards used to calculate the long term reward $T$ (in Eq. (\ref{Eq:Minimize_Cost}) and (\ref{Eq:Optimal_Policy})) is set to 600. To balance the exploration and exploitation, we would gradually increase  $\varepsilon$ from 0.5 to 0.999 with the step-size $\frac{0.499}{T_{\max}}$. All simulation results are obtained by averaging over $10^3$ independent runs.

\begin{figure} [!t]
\centering
\leavevmode \epsfxsize=2.8in  \epsfbox{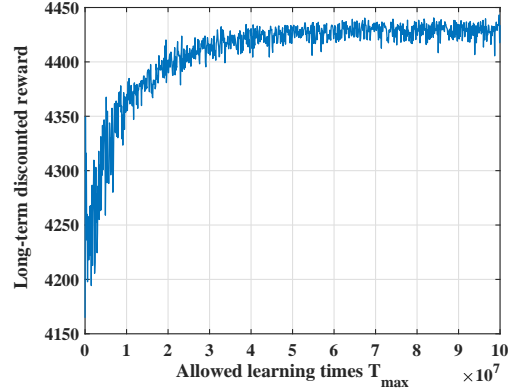}
\centering \caption{Convergence of our proposed learning algorithm EAU with $\beta_1 = \beta_2 =1$ and energy consumption $E$ evaluated in units of mJ.} \label{Fig:Fig2}
\end{figure}

Before delving into the performance of the learning algorithm EAU, we first investigate its convergence behavior by setting the allowed learning times $T_{\max}$ to different values, as shown in Fig. (\ref{Fig:Fig2}). Here, for the concerned objective function shown in Eq. (\ref{Eq:Minimize_Cost}), the energy consumption $E$ is evaluated in units of mJ, and the parameter $\beta_1$ and $\beta_2$ are both set to 1, i.e., the AoI and energy consumption are equally treated. It can be seen that our algorithm converges when $T_{\max}$ is larger than $5*10^7$. We note that the low convergence rate is resulted from the large number of state-action pairs, which is $\left| {\mathbb S \times \mathbb A} \right| = 3.2*10^5$. Besides, the fluctuations is because of the randomness in both users' behaviors and update transmission failures.

\begin{figure} [!t]
\centering
\leavevmode \epsfxsize=2.9in  \epsfbox{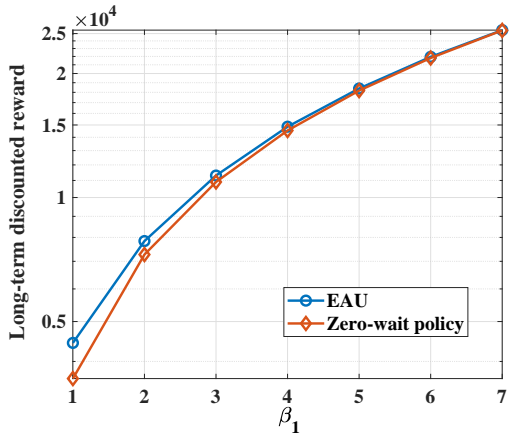}
\centering \caption{Performance comparison in terms of the long-term discounted reward, where the parameter $\beta_1$ varies from 1 to 7.} \label{Fig:Fig3}
\end{figure}

\begin{figure} [!t]
\centering
\leavevmode \epsfxsize=2.9in  \epsfbox{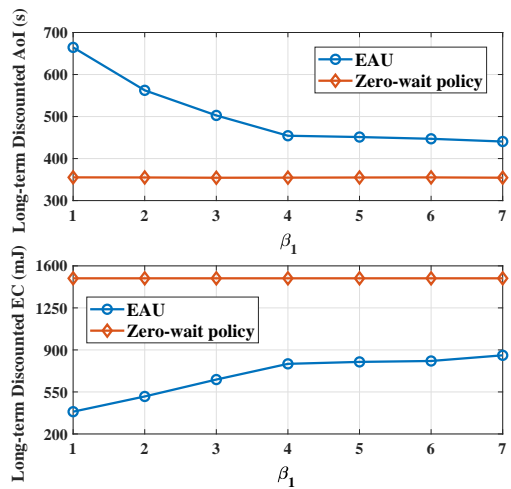}
\centering \caption{Performance comparison in terms of the long-term discounted average AoI and energy consumption, where the parameter $\beta_1$ varies from 1 to 7.} \label{Fig:Fig4}
\end{figure}

To evaluate the effectiveness of our posed algorithm, we compare its performance with the zero-wait baseline policy, i.e., the sensor updates its status in every slot, with which the AoI values at both the ECN and users would be minimized. The simulation results are demonstrated in Fig. \ref{Fig:Fig3}, where the parameter $\beta_1$ varies from 1 to 7. It should be noted that when the value of $\beta_1$ is small, our proposed algorithm EAU perform much better than the zero-wait baseline policy in terms of the achieved long-term discounted reward. However, the improvement gradually becomes less obvious as $\beta_1$ increases. This is because that when $\beta_1$ is larger, as shown in Eq. (\ref{Eq:Totoal_Cost}), reducing the weighted AoI would be more essential to lower the overall cost. To show this, we further presented the long-term discounted average AoI and energy consumption (EC) achieved by EAU and the zero-wait policy in Fig. \ref{Fig:Fig4}. It can be seen that, implementing EAU, the achieved long-term discounted average AoI and long-term discounted average EC would respectively decrease and increase, when the effect of AoI on the overall cost becomes more pronounced, i.e., $\beta_1$ is larger. However, the performance gap between EAU and the zero-wait policy becomes smaller as $\beta_1$ is larger.

\section{Conclusions}
This work presents a fresh view, by evaluating the AoI at users, for effectively striking the balance between the average AoI experienced by users and energy consumed by the sensor in caching enabled IoT networks. We formulated a dynamic status update optimization problem to minimize the expectation of a long term accumulative cost, and developed a model-free reinforcement learning algorithm to solve it. We have shown that, compared with the zero-wait policy minimizing the AoI value at both the ECN and users, our proposed scheme could achieve a higher or equal long-term discounted reward in different cases.

Based on this work, several extensions are possible. For instance, in the more realistic scenario with a large number of sensors, an interesting problem is how to design the dynamic status updating policy facing the state space and action space with extremely large sizes. Another future direction is to consider the scenario where there are correlations among the status updates from different sensors. Then, to design an efficient status update strategy for sensors, it is essential to simultaneously consider their sensing correlations as well as transmission interactions.


\bibliographystyle{IEEEtran}
\bibliography{IEEEabrv,RL_UP_Ref}

\end{document}